\documentclass[aps,prl,twocolumn,superscriptaddress]{revtex4}
\usepackage{graphicx}
\usepackage{latexsym}
\usepackage{amssymb}
\usepackage{amsmath}
\usepackage{amsfonts}
\usepackage{bm}
\usepackage{multirow}
\usepackage{color}
\newcommand{\ii}{\mathrm{i}}
\newcommand{\dd}{\mathrm{d}}
\newcommand{\vol}{\mathop{\mathrm{vol}}}

\newcommand{\dsZ}{\mathbb{Z}}
\newcommand{\Tr}{\mathop{\mathrm{Tr}}}

\newcommand{\vect}[1]{{\bm{#1}}}
\newcommand{\eqnref}[1]{Eq.\,\eqref{#1}}

\newcommand{\mat}[1]{\left(\begin{smallmatrix}#1\end{smallmatrix}\right)}
\newcommand{\beq}{\begin{equation}}
\newcommand{\eeq}{\end{equation}}
\newcommand{\beqn}{\begin{eqnarray}}
\newcommand{\eeqn}{\end{eqnarray}}

\begin{document}

\title{Bosonic Short Range Entangled states Beyond Group Cohomology classification}



\author{Cenke Xu}

\author{Yi-Zhuang You}%


\affiliation{Department of physics, University of California,
Santa Barbara, CA 93106, USA}

\begin{abstract}

We explore and construct a class of {\it bosonic} short range
entangled (BSRE) states in all $4k+2$ spatial dimensions, which
are higher dimensional generalizations of the well-known Kitaev's
$E_8$ state in $2d$~\cite{e8,kitaev_talk2}. These BSRE states
share the following properties: (1) their bulk is fully gapped and
nondegenerate; (2) their $(4k+1)d$ boundary is described by a
``self-dual" rank-$2k$ antisymmetric tensor gauge field, and it is
guaranteed to be gapless without assuming any symmetry; (3) their
$(4k+1)d$ boundary has intrinsic gravitational anomaly once
coupled to the gravitational field; (4) their bulk is described by
an effective Chern-Simons field theory with rank-$(2k+1)$
antisymmetric tensor fields, whose $K^{IJ}$ matrix is identical to
that of the $E_8$ state in $2d$; (5) The existence of these BSRE
states lead to various bosonic symmetry protected topological
(BSPT) states as their descendants in other dimensions; (6) These
BSRE states can be constructed by confining fermionic degrees of
freedom from $8$ copies of $(4k+2)d$ SRE states with fermionic
$2k-$branes; (7) After compactifying the $(4k+2)d$ BSRE state on a
closed $4k$ dimensional manifold, depending on the topology of the
compact $4k$ manifold, the system could reduce to nontrivial $2d$
BSRE states.

\end{abstract}


\maketitle

\section{1. Introduction}


In the last few years, successful classification of bosonic
symmetry protected topological (BSPT) states~\footnote{Sometimes
this kind of states are also called ``symmetry protected trivial"
states in literature, depending on the taste and level of
terminological rigor of authors.} have significantly enriched our
understanding of states of matter under strong quantum
fluctuations~\cite{wenspt,wenspt2}. The most significant property
of a SPT state is the contrast between its bulk and boundary:
while the bulk is completely gapped and nondegenerate, its
boundary remains nontrivial (gapless or degenerate) as long as the
entire system preserves certain symmetry. Meanwhile, for a BSPT
state in $d$-dimensional space, its $(d-1)$-dimensional boundary
cannot be realized (regularized) as a $(d-1)$-dimensional system
itself. The original BSPT classification~\cite{wenspt,wenspt2} was
based purely on the symmetry and dimensionality of the system, and
the symmetry group cohomology generates a list of BSPT states.

However, it is known that in both $2d$ and $3d$
spaces~\footnote{In order to avoid possible confusion, in this
paper the term ``space" always refers to real physical space
instead of space-time.}, there is one special bosonic state that
is beyond the classification table given in
Ref.~\onlinecite{wenspt,wenspt2}. In $2d$ space, this state is
usually called the ``$E_8$" state which was first proposed by
Kitaev~\cite{e8,kitaev_talk2}. The $1d$ boundary of this bosonic
state has gapless boundary state which is described by a chiral
conformal field theory with chiral central charge $c = 8$. This
$2d$ bosonic state should belong to a more general concept called
bosonic short range entangled (BSRE) state as it does not need any
symmetry to protect its boundary states. This state is most
conveniently described by a Chern-Simons (CS) theory in $(2+1)d$
space-time~\cite{luashvin}: \beqn \mathcal{S} = \int \ K^{IJ}
\frac{i}{4\pi} a_I \wedge d a_J \eeqn with an $8 \times 8$ $K$
matrix, which is precisely the Cartan matrix of the $E_8$ group.
In $3d$ space, a descendant of this state was
discovered~\cite{senthilashvin}. It is constructed as a BSPT state
protected by the time-reversal symmetry $\mathcal{T}$, and naively
it can be viewed as a state in $3d$ space with fluctuating two
dimensional $\mathcal{T}$-breaking domain walls with a $E_8$ state
confined in each domain wall. The effective field theory for this
state is $\mathcal{S} = \int \frac{i}{8\pi} K^{IJ} da_I \wedge
da_J$~\cite{senthilashvin}.

Both states mentioned above are beyond the group cohomology
classification, and they are also beyond another description of
BSPT states using semiclassical nonlinear sigma model field
theory~\cite{xuclass}. Now an obvious question is, do these BSRE
states beyond group cohomology exist only in $2d$ and $3d$, or can
they be generalized to higher spatial dimensions?

Let us look at the $E_8$ state in $2d$ space first. Unlike all the
other $2d$ BSPT states, the boundary of the $E_8$ state is chiral,
namely it only has right moving modes but no left moving modes.
This is why it does not need any symmetry to protect its boundary
states, because a chiral mode can never be back-scattered. All the
other $2d$ BSPT states have nonchiral boundary states, then their
boundary state is stable only because the left and right moving
modes carry different symmetry quantum numbers, thus the back
scattering is forbidden only with the presence of symmetry. Thus
to generalize the $E_8$ state to higher dimensions, we need to
look for higher dimensional BSRE states whose boundary is purely
``chiral".

Another way of making the same statement is that, unlike all the
other $2d$ BSPT states, the $E_8$ state boundary has intrinsic
gravitational anomaly due to its chiral nature, namely after
coupling to the gravitational field, the general coordinate
transformation at the boundary space-time is no longer a symmetry,
or in other words, the energy-momentum tensor is no longer
conserved with a background gravitational
field~\cite{wittenanomaly}. Thus to generalize the $E_8$ state to
higher dimensions, we should look for higher dimensional bosonic
states whose boundary has gravitational anomalies. Because
according to the anomaly matching
condition~\cite{anomalymatching,anomalymatching2}, if a gapless
system has perturbative gauge anomalies after coupling to a gauge
field, then it cannot be gapped out by any perturbation preserving
the gauge symmetry.

Unlike bosonic states, free fermions without interaction can
already form nontrivial SPT states. The free fermion SPT states
have been well understood and classified in
Ref.~\onlinecite{ludwigclass1,ludwigclass2,kitaevclass}, and
recent studies suggest that interaction may not lead to new SRE
states, but it can reduce the classification of fermionic SRE
states~\cite{fidkowski1,fidkowski2,qiz8,zhangz8,levinguz8,yaoz8,chenhe3B,senthilhe3,youinversion}.
It remains an open question that whether interaction can lead to
any new FSPT or FSRE state that is intrinsically different from
any free fermion state.

In this paper we will demonstrate that in all $(4k+2)d$ space,
there exists a BSRE state with stable gapless boundary without
assuming any symmetry. The existence of this BSRE state in
$(4k+2)d$ space generates a list of bosonic SPT states in higher
dimensions as its descendants. We also argue that in all $(4k+2)d$
space, there exists a SRE state constructed with ``fermionic"
$2k-$branes, $e.g.$ $2k$ dimensional objects with fermionic
statistics under exchange.

\section{2. Field Theory for SRE states in $(4k+2)d$ space}


Ref.~\onlinecite{wittenanomaly} already demonstrated that in
$(4k+1)d$ space ($(4k+2)d$ space-time), a ``self-dual" rank-$2k$
antisymmetric gauge field $\Theta_{\mu_1 \cdots \mu_{2k}}$ has
gravitational anomaly, namely the general coordinate
transformation is no longer a symmetry of the system. The
self-dual condition in the $(4k+2)d$ space-time is \beqn F = \star
F, \eeqn where $F = d\Theta$. When $k = 0$, $\Theta$ becomes a
scalar boson $\theta$, and this self-dual condition reduces to the
familiar chiral condition in $(1+1)d$ space-time: $\partial_t
\theta = \partial_x \theta$.

The chiral boson in $(1+1)d$ space-time is the boundary of a CS
field theory in $(2+1)d$ space-time. In fact, it is
straightforward to prove that a self-dual boson field in $(4k+2)d$
space-time is the boundary of a CS field theory in $(4k+3)d$
space-time with the following action~\cite{mcgreevy2013}: \beqn
\mathcal{S}_C = \int \frac{i K^{IJ}}{4\pi} C^I \wedge d C^J,
\label{actionC}\eeqn where $C^I$ is a $(2k+1)$-form antisymmetric
tensor field. At the boundary, the relation between $C$ and
$\Theta$ is $C^I = d\Theta^I$. Thus we propose that the field
theory Eq.~\ref{actionC} with a proper choice of the matrix
$K^{IJ}$ is the field theory that describes the desired BSRE
states.

The action Eq.~\ref{actionC} with $k=1$ ($6d$ space) and the
simplest choice of the matrix $K = 1$ were studied in
Ref.~\onlinecite{witten1996,mcgreevy2013}, and the authors
demonstrated that the boundary of the system must be nontrivial.
In our paper we will argue that this action with $K = 1$
corresponds to a $(4k+2)d$ SRE state constructed with fermionic
branes; while with the same $K$ matrix as the $2d$ $E_8$ state,
Eq.~\ref{actionC} corresponds to a $(4k+2)d$ BSRE state that is
beyond the group cohomology classification.

If Eq.~\ref{actionC} describes a SRE state, then there are
constraints on $K$. With $k = 0$ ($2d$ space), Eq.~\ref{actionC}
usually has topological order and topological degeneracy. On a
$2d$ torus, the ground state degeneracy (GSD) of Eq.~\ref{actionC}
is $\mathrm{Det}[K]$. On a $(4k+2)d$ torus, the GSD of
Eq.~\ref{actionC} is (please see appendix for derivation) \beqn
\mathrm{GSD} = (\mathrm{Det}[K])^{\frac{(4k+2)!}{2[(2k+1)!]^2}}.
\eeqn Since the desired BSRE and FSRE states should have no bulk
topological order or topological degeneracy, then it must have
$\mathrm{Det}[K] = 1$.

If Eq.~\ref{actionC} describes a nontrivial SRE state with nonzero
boundary gravitational anomalies, then after diagonalizing the $K$
matrix, the number of positive eigenvlues $n_+$ must be different
from the number of negative eigenvalues $n_-$, $i.e.$ $n_+ \neq
n_-$. For the simplest case $k = 0$, this implies that the
$(1+1)d$ boundary has a chiral central charge; for larger $k$,
this implies that the gravitational anomalies at the boundary do
not cancel out.

If Eq.~\ref{actionC} describes a BSRE state, then the $K$ matrix
not only needs to satisfy the previous two conditions, it also
needs to yield bosonic statistics between all excitations, which
imposes further constraints on the $K$ matrix.

\section{3. Statistics between $2k$-branes}

The ``matter field" that directly couples to the $(2k+1)$-form
gauge field is a $(2k+1)$-form current $J$, which corresponds to
the motion of a $2k$-dimensional matter ($2k$-brane) in the
space-time: \beqn \mathcal{S} &=& \mathcal{S}_{C} + \int \sum_I
m_I C^I \wedge \star J. \eeqn $m=(m_1, \cdots )^T$ is the charge
vector carried by the $2k$-brane, whose components $m_I\in\dsZ$
are all integers. Thus the simplest microscopic construction for
our SRE states is a construction based on membranes, rather than
point particles.


Suppose we consider two closed current configurations
$\mathcal{J}^{A}$ and $\mathcal{J}^{B}$ (both are closed
$(2k+1)$-manifolds) in the space-time with charge vectors $m_A$
and $m_B$ respectively, then after integrating out the gauge field
$C$, the following action will be generated: \beqn \mathcal{S} =
\int 2 \pi i \left( m_A^T K^{-1} m_B \right) J^A \wedge
M^B,\label{eq: S of J^M} \eeqn where $M^B$ is a $(2k+2)$-form
field in the space-time satisfying $\star\dd{\star M^B} = J^B$,
and the current field configuration is given by
$J^{A(B)}=\delta(x\in\mathcal{J}^{A(B)})\vol(\mathcal{J}^{A(B)})$
with $\vol(\mathcal{J})$ being the volume form of the manifold
$\mathcal{J}$. The field $M^B$ actually describes an open
$(2k+2)$-manifold $\mathcal{M}^B$ bordered by $\mathcal{J}^B$,
\emph{i.e.} $\mathcal{J}^B=\partial \mathcal{M}^B$. The non-zero
contribution to the action in \eqnref{eq: S of J^M} only comes
from the intersection of $\mathcal{J}^A$ and $\mathcal{M}^B$,
$i.e.$ when $\mathcal{J}^A$ and $\mathcal{J}^B$ are linked in the
space-time. The linking of $\mathcal{J}^A$ and $\mathcal{J}^B$ can
be interpreted as the braiding of the $2k$-branes labeled by the
charge vectors $m_A$ and $m_B$. This kind of braiding is
well-defined in arbitrary $D$-dimensional space, between two
branes of $D_1$ and $D_2$ dimensions, as long as $D = D_1 + D_2 +
2$. For instance, in $2d$ space, we can discuss braiding between
two $0d$ point particles; in $3d$ space, we can discuss braiding
between a $1d$ loop and a $0d$ point particle; and here in
$(4k+2)d$ space, \eqnref{eq: S of J^M} simply describes the
braiding statistics between two $2k$-branes.

If we take the simplest $K$ matrix, $i.e.$ $K = 1$, then two
currents $J^A$ and $J^B$ with charge $+1$ after a full braiding
will acquire phase $2\pi$, $i.e.$ two identical $2k$-branes after
exchange (half-braiding) would acquire phase $\pi$. Thus the
simplest choice of $K = 1$ corresponds to a fermionic membrane
theory.

For a bosonic state without bulk topological order or
fractionalization, all the excitations must have bosonic
statistics. Thus if we want to construct a BSRE state, we must
demand all the $2k$-branes have bosonic exchange statistics. The
simplest $K$ matrix that satisfies this criterion, while
simultaneously satisfying $\mathrm{Det}[K] = 1$ and $n_+ \neq
n_-$, is the same $K$ matrix as the $E_8$ state in $2d$ space:
\beqn K_{E_8} = \left(
\begin{array}{cccccccc}
 2 & -1 & 0 & 0 & 0 & 0 & 0 & 0 \\
 -1 & 2 & -1 & 0 & 0 & 0 & 0 & 0 \\
 0 & -1 & 2 & -1 & 0 & 0 & 0 & 0 \\
 0 & 0 & -1 & 2 & -1 & 0 & 0 & 0 \\
 0 & 0 & 0 & -1 & 2 & -1 & 0 & -1 \\
 0 & 0 & 0 & 0 & -1 & 2 & -1 & 0 \\
 0 & 0 & 0 & 0 & 0 & -1 & 2 & 0 \\
 0 & 0 & 0 & 0 & -1 & 0 & 0 & 2 \\
\end{array}
\right) \label{K}\eeqn

A bosonic brane can be constructed with point bosons on a lattice
by imposing local constraints on all the allowed configurations of
bosons, just like string configurations can be constructed with
ordinary local bosons and spins~\cite{wenstring}. Thus a bosonic
membrane model can be viewed as a local quantum boson model as
well. However, it is less obvious how to construct fermionic
branes based on local fermions. Thus the exact connection between
the fermionic membranes discussed here and local point fermion
particles requires further exploration.

The simplest fermionic SRE state in $6d$ is the integer quantum
Hall (IQH) state, whose boundary is a $5d$ chiral fermion. The IQH
state is also two copies of a $6d$ topological superconductor in
the so-called
$D$-class~\cite{ludwigclass1,ludwigclass2,kitaevclass}. We believe
that our fermionic membrane state with $K = 1$, though it is
unclear if it can be constructed with point fermion particles,
must be fundamentally different from these free fermion SRE
states. The reason is that, as was shown in
Ref.~\onlinecite{witten1984}, at the boundary of $(4k+2)d$ space,
the gravitational anomalies have multiple terms for $k > 0$. For
example, for $k = 1$, under the general coordinate transformation
$x^i \rightarrow x^i + \eta^i(x)$ at the $(5+1)d$ boundary
space-time, the gravitational anomalies have two terms: \beqn
\eta^i D^j T_{ij} \sim D^a \eta^b \epsilon_{ijklmn} \times \cr\cr
(\alpha R_{abij} R_{cdkl} R_{cdmn} + \beta R_{bcij} R_{cdkl}
R_{damn}). \eeqn The boundary of the $(4k+2)d$ IQH state is the
$(4k+1)d$ chiral fermion, whose gravitational anomaly is {\it
linearly independent} of that of the self-dual tensor field
$\Theta$ in $(4k+2)d$ space-time except for the special case
$k=0$. This implies that for $d \geq 6$, the free fermion SRE
states, and the fermionic brane state described by
Eq.~\ref{actionC} with $K = 1$ cannot be connected to each other
by deforming the boundary Hamiltonian, they must be separated by a
bulk quantum phase transition.

\section{4. Descendant BSPT states in other dimensions}


Once we identify a SRE state in $(4k+2)d$ space, we can construct
a series of SPT states in other dimensions as its descendants.
Unlike their parent SRE state in $(4k+2)d$ space, these SPT states
do need certain symmetry to protect their boundary
states.\author{Cenke Xu}

One such BSPT state in $3d$ space was constructed in
Ref.~\onlinecite{senthilashvin} as a descendant of the $2d$ $E_8$
state. This $3d$ BSPT state is protected by the time-reversal
symmetry $\mathcal{T}$. This BSPT state is constructed as follows:
we first break $\mathcal{T}$, then the system will become a
trivial state. Then $\mathcal{T}$ can be restored by proliferating
the domain walls between $\mathcal{T}$-breaking domains. It is the
domain walls that distinguish this $3d$ BSPT state from other
trivial $3d$ states: there is a $2d$ $E_8$ state sandwiched in
each $2d$ $\mathcal{T}$-breaking domain wall. Also, the ordinary
$3d$ topological insulator can be viewed as a $3d$ state with
proliferating $\mathcal{T}$-breaking domain walls with a $2d$ $\nu
= 1$ IQH state sandwiched in each domain wall.

The same construction can be trivially generalized to any
$(4k+3)d$ space: a $(4k+3)d$ BSPT state can be constructed by
proliferating domain walls of $\mathcal{T}$-breaking domains, with
the $(4k+2)d$ BSRE state sandwiched in each domain wall. The
effective field theory for such $(4k+3)d$ state is \beqn
\mathcal{S} = \int \ \frac{i\theta}{2\pi} \frac{1}{4\pi} K^{IJ} d
C^I \wedge d C^J. \label{4k+3}\eeqn Analogous to the simplest case
with $k = 0$ discussed in Ref.~\onlinecite{senthilashvin}, and
similar to the ordinary $3d$ topological insulator~\cite{qi2008},
when $\theta = \pi$ this state is a nontrivial $(4k+3)d$ BSPT with
time-reversal symmetry, while when $\theta = 0$ mod $2\pi$, this
state is trivial. Thus this $(4k+3)d$ BSPT has $Z_2$
classification.

The $Z_2$ classification can be understood as the follows. On a
closed $(4k+4)d$ space-time manifold, the integral $ \int
\frac{1}{4\pi^2} dC \wedge dC $ is an even integer on any manifold
whose $(2k+2)$th Wu-class vanishes, while it is an odd integer
otherwise~\cite{witten1996,moore2006}. When $k=0$, the second
Wu-class is equivalent to the second Stiefel-Whitney
class~\cite{moore2006}, thus the condition of vanishing Wu-class
reduces to the well-known condition for a manifold to allow a spin
structure. However, with the BSRE $K$ matrix in Eq.~\ref{K}, the
integral $\int \frac{1}{4\pi^2} K^{IJ} d C^I \wedge d C^J$ is
always an even integer on any closed $(4k+4)d$ space-time
manifold. This implies that in Eq.~\ref{4k+3} $\theta$ and $\theta
+ 2\pi$ are equivalent. And because time-reversal transformation
takes $\theta$ to $ - \theta$, if the system has time-reversal
symmetry, it demands $\theta$ take only discrete values $\theta =
\pi n$. While with $\theta = 2\pi$, if $\mathcal{T}$ is broken at
the boundary, the $(4k+2)d$ boundary of Eq.~\ref{4k+3} is
precisely the $(4k+2)d$ theory described in Eq.~\ref{actionC},
thus the boundary can be regularized as a $(4k+2)d$ system itself.
Thus $\theta = 2\pi$ corresponds to a trivial state. When $\theta
= \pi$, naively the boundary corresponds to ``half" of the
$(4k+2)d$ BSRE state in Eq.~\ref{actionC}, thus it cannot be
realized as a $(4k+2)d$ state itself. When $\theta = \pi$, if the
boundary preserves $\mathcal{T}$, based on our experiences with
$3d$ topological insulator, we expect the boundary to be either
gapless, or have $(4k+2)d$ topological order that cannot be
regularized on $(4k+2)d$ space itself.

Similar constructions can be generalized to BSPT states with other
symmetries as well. For every $(4k+4)d$ space, we can construct a
BSPT state with U(1) symmetry as follows: We start with the
superfluid phase with spontaneous U(1) symmetry breaking, then the
U(1) symmetry can be restored by proliferating U(1) vortices. In
$(4k+4)d$ space, a vortex is a $(4k+2)$-brane. For ordinary
systems, after proliferating the vortices the system will enter a
trivial gapped disordered phase. However, if there is a $(4k+2)d$
BSRE state described by Eq.~\ref{actionC} confined in each vortex,
then after vortex proliferation the system will enter the desired
BSPT state with U(1) symmetry. Once we couple the U(1) symmetry to
an external U(1) gauge field $A_\mu$, this system is described by
the following effective field theory: \beqn \mathcal{S} = \int \
\frac{i}{8\pi^2} K^{IJ} C^I \wedge d C^J \wedge d A.
\label{4k+4}\eeqn This state has $Z$ classification, since for
arbitrary copies of this system, it remains nontrivial.

At the boundary of the $(4k+4)d$ theory Eq.~\ref{4k+4}, after
coupling to both gravitational field and U(1) gauge field, there
is a mixed gauge-gravitational anomaly, $i.e.$ the conservation of
energy momentum tensor is violated in the flux of the U(1) gauge
field. For example, for $k = 1$, at the seven dimensional boundary
(eight dimensional boundary space-time), the mixed anomaly reads:
\beqn \eta^i D^j T_{ij} \sim D^a \eta^b \epsilon_{ijklmn pq}
\times \cr\cr (\alpha R_{abij} R_{cdkl} R_{cdmn} + \beta R_{bcij}
R_{cdkl} R_{damn})
\partial_p A_q. \eeqn

For every $(4k+2+2n)d$ space, we can construct a family of BSPT
state with U(1) symmetry, based on the BSRE state in $(4k+2)d$.
This state is constructed as follows: In $(4k+2+2n)d$ space, a
U(1) vortex is a $(4k+2n)$ dimensional membrane. Then $n$ vortices
will intersect on a $(4k+2)$ dimensional membrane, where the
$(4k+2)d$ BSRE state Eq.~\ref{actionC} can reside. Then this BSPT
state discussed in this paragraph can be obtained by proliferating
these vortices with BSRE state Eq.~\ref{actionC} confined in each
$n-$vortex intersection. After coupling the system to an external
U(1) gauge field, the effective field theory will be \beqn
\mathcal{S} = \int \ \frac{i}{4\pi (2\pi)^n n!} K^{IJ} C^I \wedge
d C^J \wedge (d A)^n. \label{4k+2+2n}\eeqn

Again, at the $(4k+2n+1)d$ boundary of Eq.~\ref{4k+2+2n}, after
coupling to both gravitational field and U(1) gauge field, there
is a mixed gauge-gravitational anomaly. All these BSPT states
constructed in this section are beyond the group cohomology
classification developed in Ref.~\onlinecite{wenspt,wenspt2}.

\section{5. Confine fermionic branes}

In this section we will demonstrate that the BSRE state with $K =
K_{E_8}$ in Eq.~\ref{K} can be constructed with fermionic brane
theories with $K = 1$.

We start from 8 copies of $K=1$ theory, which is still a CS field
theory but with the $K$ matrix being an $8\times 8$ identity
matrix, denoted as $I_8$. To make connection to the $E_8$ theory,
we first attach a trivial bosonic state to the system, which is
described by the $K$ matrix $\sigma^x$. Thus the entire system has
$K$ matrix $I_8 \oplus \sigma^x$. $\sigma^x$ has eigenvalues $+1$
and $-1$, thus if a system has $K =\sigma^x $, its boundary has
both self-dual and anti self-dual fields, $i.e.$ its boundary has
no gravitational anomaly, hence it is a trivial bosonic state in
$(4k+2)d$.

Just like $2d$ Abelian quantum Hall state, different $K$ matrices
can correspond to the same physical state, as long as these $K$
matrices only differ from each other by a GL$(N, \mathbb{Z})$
transformation $W$ ($\text{det}\,W=1$)~\cite{wenzee}. The physical
meaning of the GL$(N, \mathbb{Z})$ transformation is just to
relabel all the excitations. It is straight forward to verify that
there exists the following transformation $W$,
\begin{equation}
W=\left(
\begin{smallmatrix}
 1 & 0 & 0 & 0 & 0 & 0 & 0 & -1 & -1 & 2 \\
 -1 & 1 & 0 & 0 & 0 & 0 & 0 & -1 & -1 & 2 \\
 0 & -1 & 1 & 0 & 0 & 0 & 0 & -1 & -1 & 2 \\
 0 & 0 & -1 & 1 & 0 & 0 & 0 & -1 & -1 & 2 \\
 0 & 0 & 0 & -1 & 1 & 0 & 0 & -1 & -1 & 2 \\
 0 & 0 & 0 & 0 & -1 & 1 & 0 & 0 & -1 & 2 \\
 0 & 0 & 0 & 0 & 0 & -1 & 1 & 0 & -1 & 2 \\
 0 & 0 & 0 & 0 & 0 & 0 & -1 & 1 & 0 & -1 \\
 0 & 0 & 0 & 0 & 0 & 0 & 0 & 1 & 1 & -3 \\
 0 & 0 & 0 & 0 & 0 & 0 & 1 & -2 & -3 & 5 \\
\end{smallmatrix}
\right),
\end{equation}
such that \beqn W^\intercal (I_8\oplus\sigma^x) W =
K_{E_8}\oplus\sigma^z. \eeqn Existence of this transformation
implies that 8 copies of fermionic membrane theories plus a
trivial boson state, is topologically equivalent to the desired
$(4k+2)d$ BSRE state attached to a trivial fermionic brane state
with $K = \sigma^z$ (the boundary of Eq.~\ref{actionC} with $K =
\sigma^z$ is also nonchiral, hence it corresponds to a trivial
state).

Since eventually the $K$ matrix $K_{E_8}\oplus\sigma^z$ is block
diagonal, fermionic and bosonic degrees of freedom are decoupled.
This implies that we can safely tune the gap of fermionic branes
to infinity without interfering with the bosonic sector. Thus 8
copies of fermionic brane theory with $K = 1$ will automatically
become the desired BSRE state with $K = K_{E_8}$ given by
Eq.~\ref{K}, after the fermionic branes are confined.

\section{6. Reduction to $2d$ space}

A natural question to ask about the SRE states constructed by
Eq.~\ref{actionC} is that, if we define the system on a $(4k+2)d$
space manifold that is $\mathbb{R}^2 \otimes \mathcal{M}_{4k}$,
where $\mathbb{R}^2$ is the infinite two dimensional plane, while
$\mathcal{M}_{4k}$ is a closed compact $4k-$dimensional manifold,
then does this state reduce to any nontrivial SRE state in $2d$?

Let us take $k = 1$ ($6d$ space with $5d$ boundary) and $K = 1$ as
an example. At the $5d$ boundary, we can choose a gauge to make
$\Theta_{0j} = 0$. Then the self-dual condition reads \beqn
\partial_t \Theta_{ij} = \epsilon_{ijklm} \partial_k \Theta_{lm}.
\eeqn The dynamical modes in the $(1+1)d$ space-time $(t, x_1)$
are $\Theta_{ab}$ with $a, b = 2 \cdots 5$: \beqn
\partial_t\Theta_{23} &=&  \partial_1 \Theta_{45} +
\partial_4 \Theta_{51} + \partial_5 \Theta_{14};  \cr\cr
\partial_t\Theta_{45} &=&  \partial_1 \Theta_{23} +
\partial_2 \Theta_{31} + \partial_3 \Theta_{12}; \cr\cr \cdots
& = & \cdots  \eeqn

Let us define a $1-$form vector field $\vec{A} = (\Theta_{12},
\Theta_{13}, \Theta_{14}, \Theta_{15})$ on the compact
$4-$manifold $\mathcal{M}_4$, then the chirality of the reduced
field theory on the remaining $(1+1)d$ space-time, depends on
whether the corresponding theory on $\mathcal{M}_{4}$ is self-dual
or anti self-dual. For example, if $d_4 A = \star d_4 A$
(self-dual) on $\mathcal{M}_4$, then $\Theta_{23} + \Theta_{45}$,
$\Theta_{34} + \Theta_{25}$, $\Theta_{35} + \Theta_{42}$ are three
left moving modes; while if $d_4 A = - \star d_4 A$ (anti
self-dual) on $\mathcal{M}_4$, then $\Theta_{23} - \Theta_{45}$,
$\Theta_{34} - \Theta_{25}$, $\Theta_{35} - \Theta_{42}$ are three
right moving modes. Each self-dual (anti self-dual) configuration
$\vec{A}$ on $\mathcal{M}_4$ will produce three left (right)
moving modes on the reduced $(1+1)d$ space-time.

Thus with $K = 1$, the chiral central charge of the reduced
$(1+1)d$ theory is \beqn c = 3 (n_+ - n_-), \eeqn where $n_\pm$ is
the number of independent 2-form $\omega_{\pm}$ on $\mathcal{M}_4$
that satisfies \beqn && \mathbf{d} \omega_+ = 0, \ \ \ \omega_+ =
\star \omega_+,\cr\cr && \mathbf{d} \omega_- = 0, \ \ \ \omega_- =
- \star \omega_-, \eeqn where $\mathbf{d} = d - \star d \star$.
Mathematically $\omega_+$ and $\omega_-$ are the kernel and
cokernal of operator $\mathbf{d}$, and $n_+ - n_-$ is the index of
operator $\mathbf{d}$, and according to the Atiyah-Singer theorem,
the index is a topological invariant of $\mathcal{M}_4$. In this
particular case $n_+ - n_-$ corresponds to the signature (also
called the $L-$genus) of $\mathcal{M}_4$: $n_+ - n_- =
\mathrm{Sign}[\mathcal{M}_4]$. The signature of a manifold is
determined by its topology only. The signature of some example
$4-$manifolds is listed in the follows~\cite{TFTbook}: \beqn
\mathrm{Sign}[S^4] = 0, \ \ \ \mathrm{Sign}[T^4] = 0, \ \ \
\mathrm{Sign}[\mathrm{CP}^2] = 1. \eeqn

On $S^4$ and $T^4$, because their signature is zero, after
compactification the reduced $(1+1)d$ theory is nonchiral, thus
the reduced $2d$ bulk theory is a trivial disordered state.
However because $\mathrm{Sign}[\mathrm{CP}^2] = 1$, if we take $K
= 1$ in Eq.~\ref{actionC}, then after compatifying
Eq.~\ref{actionC} on CP$^2$, the reduced theory on $(1+1)d$
space-time has chiral central charge $3$, thus the $2d$ bulk is
presumably equivalent to $\nu = 3$ integer quantum Hall state in
$2d$. If we take $K = K_{E_8}$ in Eq.~\ref{K}, then the reduced
$2d$ bulk becomes a nontrivial BSRE state in $2d$, which
presumably is equivalent to three copies of the $2d$ $E_8$ states.

\section{7. Summary and Discussion}



In this work we construct BSRE states in $(4k+2)d$ space, along
with its descendant BSPT states in other dimensions, using the
fact that the boundary of Eq.~\ref{actionC} has intrinsic
gravitational anomalies after coupling to the gravitational field.
Due to the anomaly matching condition, the gravitational anomalies
imply that the boundary of Eq.~\ref{actionC} cannot be gapped out,
even though we assume no symmetry at all in Eq.~\ref{actionC}.

So far the gravitational anomalies we used in our work were all
perturbative gravitational anomalies. There is another type of
global gravitational anomalies that we have not discussed yet. In
Ref.~\onlinecite{witten1984}, it was demonstrated that a single
(or odd flavors of) Majorana fermion in $8k-1$ and $8k$
dimensional space has global gravitational anomalies, namely the
partition function of the system changes sign under a large global
coordinate transformation. These global gravitational anomalies
have $Z_2$ classification, because even flavors of Majorana
fermions in these dimensions have no global gravitational
anomalies. These global anomalies precisely correspond to the
$Z_2$ classification of topological superconductors in $8k$ and
$8k+1$ dimensional space without any symmetry (the so called $D$
class~\cite{ludwigclass1,ludwigclass2,kitaevclass}), whose
boundary is precisely massless Majorana fermion in $8k-1$ and $8k$
dimensional space.

Ref.~\onlinecite{kapustin1,kapustin2,kapustin3,kapustin4,kongwen},
proposed that there exists a $4d$ BSRE state without any symmetry,
and this state has $Z_2$ classification. This result implies that
the $3d$ boundary of odd flavors of this BSRE state should have
global gravitational anomalies. Ref.~\onlinecite{mcgreevy2014}
proposed that the effective field theory description of this $4d$
BSRE state is a $(4+1)d$ version of Eq.~\ref{actionC}, and the
gauge field involved is a $2-$form field. Also
Ref.~\onlinecite{mcgreevy2014} pointed out that this state
corresponds to the $4d$ BSRE state discussed in
Ref.~\onlinecite{wangsenthilscience} whose boundary is a $3d$ QED
with fermionic gauge charge and monopole. More general connection
between global gravitational anomaly and SRE states is desired,
which we will leave to future studies.

\section{Acknowledgement}

The authors are supported by the the David and Lucile Packard
Foundation and NSF Grant No. DMR-1151208. Authors also acknowledge
very helpful discussions with Joe Polchinski, T. Senthil,
Xiao-Gang Wen and Xi Yin. Especially the authors want to thank Xi
Yin for providing very helpful mathematical references.

\bibliography{beyond}

\maketitle \onecolumngrid
\appendix

\section{Quantization of $K$}

The $(2k+1)$-form gauge fields $C$ that we considered here are
\emph{compact} U(1) gauge fields. The compactness arise from the
assumption that the matter of $2k$-brane only carries quantized
(integer) gauge charge, \emph{i.e.} any $2k$-brane is an integer
multiple of the elementary $2k$-brane. Therefore all physical
observables in the gauge theory are given by (functions of) the
Wegner-Willson amplitudes on some closed $(2k+1)d$ current
manifold $\mathcal{J}$ with the elementary gauge charge,
\begin{equation}
\mathcal{W}[\mathcal{J}]=e^{\ii\int_\mathcal{J} C}.
\end{equation}
Any transformations of $C$ that keep all the Wegner-Willson
amplitudes $\mathcal{W}[\mathcal{J}]$ invariant are gauge
transformations, which include the local gauge transformations
$C\to C+\dd \Theta$ (induced by arbitrary $2k$-form fields
$\Theta$) that trivially preserve $\mathcal{W}[\mathcal{J}]$, as
well as the large gauge transformations which are allowed only in
the compact gauge theory. In a $(4k+3)d$ space-time with
periodicity $L_i$ in the $\dd x^i$ direction, the large gauge
transformation takes the following form
\begin{equation}\label{eq: large gauge}
C\to C+\delta C\text{, with }\delta C=\frac{2\pi N_{i_1\cdots
i_{2k+1}}}{L_{i_1}\cdots L_{i_{2k+1}}}\dd
x^{i_1}\wedge\cdots\wedge \dd x^{i_{2k+1}},
\end{equation}
with integers $N_{i_1\cdots i_{2k+1}}\in\dsZ$ for
$i_1<\cdots<i_{2k+1}$. For  \emph{contractable} current manifold
$\mathcal{J}=\partial \mathcal{M}$,
$\int_\mathcal{J}C=\int_{\partial\mathcal{M}}C=\int_{\mathcal{M}}\dd
C$ is simply invariant under the transformation in \eqnref{eq:
large gauge} as $\dd \delta C = 0$, so $\mathcal{W}[\mathcal{J}]$
is also invariant. For \emph{non-contractable} current manifold
$\mathcal{J}$, $\int_\mathcal{J}\delta C=2\pi N$ will be an
integer multiple of $2\pi$, but $\mathcal{W}[\mathcal{J}]\to
e^{\ii 2\pi N}\mathcal{W}[\mathcal{J}]=\mathcal{W}[\mathcal{J}]$
is still invariant. Hence it is verified that \eqnref{eq: large
gauge} is indeed an allowed gauge transformation.

The invariance of the partition function $\mathcal{Z}=\Tr
e^{-\mathcal{S}_C}$ with $ \mathcal{S}_C = \int \frac{\ii
K^{IJ}}{4\pi} C^I \wedge \dd C^J$ in \eqnref{actionC} under the
large gauge transformation \eqnref{eq: large gauge} necessarily
requires the entries of the $K$ matrix to be integers, \emph{i.e.}
$K^{IJ}\in\dsZ$. For simplicity, let us first take the case of
single-component gauge field $C$, and consider the large gauge
transformation $C_{0\cdots 2k} \to C_{0\cdots 2k} +
\frac{2\pi}{L_0\cdots L_{2k}}$, the action will be changed by
\begin{equation}\label{eq: S of C under large gauge}
\delta \mathcal{S}_C=\int\frac{\ii K}{4\pi}
\frac{2\times2\pi}{L_0\cdots L_{2k}}\dd x^0\wedge\cdots\wedge\dd
x^{2k}\wedge \dd C=\ii K\int_{\Omega_{2k+2}}\dd C,
\end{equation}
where $\Omega_{2k+2}$ is the closed $(2k+2)$-manifold
parameterized by the remaining coordinates $(x^{2k+1},\cdots,
x^{4k+2})$. It can be proved that $\int_{\Omega_{2k+2}}\dd C$ must
be an integer multiple of $2\pi$ on any closed $(2k+2)$-manifold
$\Omega_{2k+2}$ by generalizing the argument in
Ref.\,\onlinecite{Kohmoto}. First consider the manifold
$\Omega_{2k+2}$ as patches glued together, such that on each patch
the gauge field $C$ is smooth and single-valued, then
$\int_{\Omega_{2k+2}}\dd C=\int_{\partial\Omega_{2k+2}}C$ with
$\partial\Omega_{2k+2}$ being the boundary of the patches, on
which $C$ from both sides must be related by some gauge
transformation $C\to C+\dd \Theta$. So the integral will become
$\int_{\Omega_{2k+1}}\dd \Theta$ on one side of the patch
boundary.
Keep using the same trick on $\Omega_{2k+1}$: consider
$\Omega_{2k+1}$ as $(2k+1)$-dimensional patches glued along their
$2k$-dimensional boundary, then the integral can be ascribed to
the difference of $\Theta$ form both sides (denoted as
$\Theta_\pm$) along the boundary manifold $\Omega_{2k}$ as
$\int_{\Omega_{2k}}(\Theta_+-\Theta_-)$. According to the
compactification condition, $\int_{\Omega_{2k}}\Theta$ and $2\pi +
\int_{\Omega_{2k}}\Theta$ are equivalent on any closed
$2k$-manifold $\Omega_{2k}$, because it corresponds to changing
the phase of the matter field ($2k$ dimensional membranes) by
$2\pi$. So $\int_{\Omega_{2k}}\Theta_+$ and
$\int_{\Omega_{2k}}\Theta_-$ can only differ by multiples of
$2\pi$, and hence proved $\int_{\Omega_{2k+2}}\dd
C=\int_{\Omega_{2k}}(\Theta_+-\Theta_-)=2\pi N$ with $N\in\dsZ$.
Substituting this result back to \eqnref{eq: S of C under large
gauge}, we arrive at the conclusion that under large gauge
transformation, the action is changed by
$\delta\mathcal{S}_C=2\pi\ii K N$ with $N\in\dsZ$, so $K$ must
also be an integer to ensure the invariance of the partition
function $\mathcal{Z}=\Tr e^{-\mathcal{S}_C}$ for any $N$. It is
straight forward to generalize the above argument to
multi-component $C^I$ field, and the conclusion is that every
element $K^{IJ}\in\dsZ$ in the $K$ matrix must be an integer.

\section{ground state degeneracy of Eq.~\ref{actionC}}

The topological ground state degeneracy of the Chern-Simon theory
on the torus can be calculated following
Ref.\,\onlinecite{wenzeeCS, BarkeshliWen}. We first fix the
temporal gauge to $C_{0i_1\cdots i_{2k}}=0$ (where
$i_1,\cdots,i_{2k}$ are spacial indices) by the gauge
transformation $C_{0i_1\cdots i_{2k}}\to C_{0i_1\cdots
i_{2k}}+\partial_0 \Theta_{i_1\cdots i_{2k}}$ with properly chosen
$\Theta_{i_1\cdots i_{2k}}$ (there are totally $\mat{4k+2\\2k}$ of
$C_{0i_1\cdots i_{2k}}$ to be fixed by the same number of
$\Theta_{i_1\cdots i_{2k}}$). The equations of motion for
$C_{0i_1\cdots i_{2k}}$ act as constraints requiring $\dd C=0$ in
the $(4k+2)d$ space for the spatial components $C_{i_1\cdots
i_{2k+1}}$. This implies that the gauge-inequivalent
configurations are completely specified by the holonomies of the
spacial gauge field $C$ on the non-contractible $(2k+1)$-manifolds
$\mathcal{J}$ of the torus, \emph{i.e.} by $\int_\mathcal{J}C$ for
every homology basis $\mathcal{J}$. This configuration space can
be parameterized by an antisymmetric rank-$(2k+1)$ tensor $X$ as
\begin{equation}
C_{i_1\cdots i_{2k+1}}(t,\vect{x})=\frac{2\pi}{L_{i_1}\cdots
L_{i_{2k+1}}}X_{i_1\cdots i_{2k+1}}(t).
\end{equation}
The large gauge transformation in \eqnref{eq: large gauge} takes
$X_{i_1\cdots i_{2k+1}}\to X_{i_1\cdots i_{2k+1}}+N_{i_1\cdots
i_{2k+1}}$ with $N_{i_1\cdots i_{2k+1}}\in \dsZ$, thus
$X_{i_1\cdots i_{2k+1}}\sim X_{i_1\cdots i_{2k+1}}+1$ labels the
same quantum state, and the quantum mechanical wave function must
respect this periodicity of the configuration space. With this
parameterization, the Chern-Simons action reads
\begin{equation}
\mathcal{S}_C=\frac{2\pi\ii K}{2}\int\dd t\; X\wedge \dot{X},
\end{equation}
so the momentum conjugate to $X$ is $p_X=\delta
\ii\mathcal{S}_C/\delta \dot{X}=2\pi K {\star X}$. There are all
together
$\frac{1}{2}\mat{4k+2\\2k+1}=\frac{(4k+2)!}{2[(2k+1)!]^2}$ pairs
of canonical conjugate variables among the components of $X$. Each
pair contributes $K$-fold ground state degeneracy (GSD), as the
periodicity $X\sim X+1$ requires the momentum quantization
$p_X=2\pi K {\star X}=2\pi n$ (with $n\in\dsZ$), then $\star
X\sim\star X+1$ would imply $n\sim n+K$ so that $n=0,\cdots,K-1$
labels $K$ degenerated states. So for
$\frac{(4k+2)!}{2[(2k+1)!]^2}$ pairs of conjugate variables in
$X$, the total GSD is
\begin{equation}
\text{GSD}=K^\frac{(4k+2)!}{2[(2k+1)!]^2}.
\end{equation}
For multi-component gauge theory with a $K$ matrix, the number $K$
in the above formula is just replaced by $\det K$.

\end{document}